\documentclass[twocolumn,showpacs,preprintnumbers,amsmath,amssymb]{revtex4}

\usepackage{graphicx}
\usepackage{dcolumn}
\usepackage{bm}
\usepackage{epsfig} 
\usepackage{subfigure}

\def\Mpc{\,{\rm Mpc}}

\def\la{\mathrel{\mathpalette\fun <}}
\def\ga{\mathrel{\mathpalette\fun >}}
\def\fun#1#2{\lower3.6pt\vbox{\baselineskip0pt\lineskip.9pt
  \ialign{$\mathsurround=0pt#1\hfil##\hfil$\crcr#2\crcr\sim\crcr}}}

\input epsf

\def\plotone#1{\centering \leavevmode
\epsfxsize= 1.0\columnwidth \epsfbox{#1}}

\newenvironment{tablehere}{\def\@captype{table}}{}
\newcommand{\tableskip}{\\[-6pt]}
\def\be{\begin{equation}}
\def\ee{\end{equation}}
\def\ba{\begin{eqnarray}}
\def\ea{\end{eqnarray}}
\def\bea{\begin{eqnarray}}
\def\eea{\end{eqnarray}}

\begin{document}

\def\affilmrk#1{$^{#1}$}
\def\affilmk#1#2{$^{#1}$#2;}

\def\davis{1}
\def\ita{2}
\def\cma{3}
\def\jpl{4}
\def\cit{5}
\def\warsaw{6}
\def\jpl2{7}
\def\uiucp{8}
\def\uiuca{9}


\title{Cosmological Parameter Constraints as Derived from the
\emph{Wilkinson Microwave Anisotropy Probe} Data via Gibbs Sampling
and the Blackwell-Rao Estimator}

\author{
M. Chu\affilmrk{\davis}\footnote{email:mchu@bubba.physics.ucdavis.edu},
H. K. Eriksen\affilmrk{\ita,\cma,4,\cit}\footnote{email: h.k.k.eriksen@astro.uio.no},
L. Knox\affilmrk{\davis}\footnote{email:knox@bubba.physics.ucdavis.edu}, 
K. M. G\'{o}rski\affilmrk{4,\cit,\warsaw},
\\J. B. Jewell\affilmrk{\jpl2},
D. L. Larson\affilmrk{\uiucp},
I. J. O'Dwyer\affilmrk{\uiuca},
B. D. Wandelt\affilmrk{\uiucp,\uiuca}
}

\affiliation{
\parshape 1 -3cm 24cm
\affilmk{\davis}{Department of Physics, One Shields Avenue, University of California, Davis, California 95616}\\
\affilmk{\ita}{Institute of Theoretical Astrophysics, University of
Oslo, P.O.\ Box 1029 Blindern, N-0315 Oslo, Norway}\\
\affilmk{\cma}{Centre of
Mathematics for Applications, University of Oslo, P.O.\ Box 1053
Blindern, N-0316 Oslo, Norway}\\
\affilmk{4}{Jet Propulsion Laboratory, M/S 169/327, 4800
Oak Grove Drive, Pasadena CA 91109}\\
\affilmk{\cit}{California Institute of
Technology, Pasadena, CA 91125}\\
\affilmk{\warsaw}{Warsaw University Observatory, Aleje
Ujazdowskie 4, 00-478 Warszawa, Poland}\\
\affilmk{\jpl2}{Jet Propulsion Laboratory, M/S 126/347, 4800 Oak
Grove Drive, Pasadena CA 91109}\\
\affilmk{\uiucp}{Department of Physics, University of Illinois at Urbana-Champaign, IL 61801-3080}\\
\affilmk{\uiuca}{Department of Astronomy, University of Illinois at
  Urbana-Champaign, IL 61801-3080}
}

\date{\today}

\begin{abstract}
We study the Blackwell-Rao (BR) estimator of the probability
distribution of the angular power spectrum, $P(C_{\ell}|\mathbf{d})$,
by applying it to samples of full-sky no-noise CMB maps generated via
Gibbs sampling.  We find the estimator, given a set of samples, to be
very fast and also highly accurate, as determined by tests with
simulated data.  We also find that the number of samples required for
convergence of the BR estimate rises rapidly with increasing $\ell$,
at least at low $\ell$.  Our existing sample chains are only long
enough to achieve convergence at $\ell \la 40$.  In comparison with
$P(C_{\ell}|\mathbf{d})$ as reported by the \emph{WMAP} team we find
significant differences at these low $\ell$ values.  These differences lead
to up to $\sim 0.5 \ \sigma$ shifts in the estimates of parameters in
a 7-parameter $\Lambda$CDM model with non-zero
$\textrm{d}n_{\textrm{s}}/\textrm{d}\ln k$.  Fixing
$\textrm{d}n_{\textrm{s}}/\textrm{d}\ln k = 0$ makes these shifts much
less significant.

Unlike existing analytic approximations, the BR estimator can be
straightforwardly extended for the case of power spectra from
correlated fields, such as temperature and polarization.  We discuss
challenges to extending the procedure to higher $\ell$ and provide
some solutions.
\end{abstract}



\maketitle


\section{Introduction}

As predicted \citep{spergel95,knox95,jungman96a}, observations of the
cosmic microwave background (CMB) anisotropies
(e.g. \citep{kuo04,bennett03,readhead04}) have provided very tight
constraints on cosmological parameters
(e.g. \citep{spergel03,goldstein03,rebolo04}).  The standard approach
to estimating cosmological parameters, given a map of the CMB, 
is to first estimate the probability
distribution of the angular power spectrum from the map or
time-ordered data, $P(C_{\ell}|\mathbf{d})$, and then use
$P(C_{\ell}|\mathbf{d})$ to get the probability distribution of the
cosmological parameters assuming some model.  While it is possible to
estimate the cosmological parameters without ever estimating
$P(C_{\ell}|\mathbf{d})$, going through this intermediate step has
several advantages.  Chief among these is that one can estimate
parameters for many different parameter spaces, each time starting
from the same $P(C_{\ell}|\mathbf{d})$ instead of from an earlier
point in the analysis pipeline, thereby reducing demands on computer
resources.

The path from $P(C_{\ell}|\mathbf{d})$ to cosmological parameter
constraints is most often traversed by the generation of a Monte Carlo
Markov chain (MCMC) \cite{gilks96,christensen01,knox01b}.  The chain is a list of
locations in the cosmological parameter space which has the useful
property that the probability that the true value is in some region of
parameter space is proportional to the number of chain elements in
that region of parameter space.  The chain is generated using a
Metropolis-Hastings algorithm that requires evaluation of
$P(C_{\ell}|\mathbf{d})$ at tens of thousands of randomly chosen trial
locations.

At low $\ell$ $P(C_{\ell}|\mathbf{d})$ is significantly non-Gaussian.
Non-Gaussian analytic forms, whose parameters can be estimated from
the data, have been investigated \citep{bond00,bartlett00,verde03} and
widely used.  The validity of these analytic approximations however is
not under rigorous mathematical control.  It is established on a
case-by-case basis by comparison with computationally expensive
brute-force evaluations of $P(C_{\ell}|\mathbf{d})$.  Further, these
comparisons do show some level of discrepancy which may be significant
for parameter estimation.

Here we calculate $P(C_{\ell}|\mathbf{d})$ with the BR
estimator as suggested by \citet{wandelt03}.  This estimator is a sum
over $P(C_{\ell} |\mathbf{s}_i)$ where the $\mathbf{s}_i$ are a chain 
of possible all-sky signal maps produced as a by-product of the Gibbs 
sampling procedure.  The BR estimator has some appealing properties.  
First, it is exact in the limit of an infinite number of samples.  
Second, given the samples, it can be very rapidly calculated.

Of course, the BR estimator is only accurate given a sufficient number
of samples for convergence.  We study convergence of the BR estimate
from samples generated from first-year {\it WMAP} Q, V and W band data
as described by \citet{eriksen04} and \citet{odwyer04}.  We find that
the number of samples rises exponentially with increasing maximum
multipole considered, $\ell_{\rm max}$, due to the rising volume of
the space to be explored.  Beyond $\ell \sim 40$ we need more samples
than the 955 that we have.

Even at $\ell \le 30$ where our BR estimate has converged, we find
significant differences between our BR-estimated $P(C_{\ell}
|\mathbf{d})$ and that given by the {\it WMAP} team as described by
\citet{hinshaw03} (hereafter H03) and \citet{verde03}.  These
differences are not due solely to BR though, but the combined effect
of a number of differences in our analysis procedures.  To investigate
the significance of these differences we estimate cosmological
parameters in two cases: i) using the \emph{WMAP} team's description
of $P(C_{\ell}|\mathbf{d})$ and ii) using a hybrid scheme where we
replace the \emph{WMAP} team's $P(C_{\ell}|\mathbf{d})$ at $\ell \le
30$ with the BR estimate.  Assuming a zero mean curvature $\Lambda$CDM
cosmology parameterized by the primordial power spectrum amplitude and
power-law index, reionization redshift, baryon density, cold dark
matter density and a cosmological constant, we find no significant
changes to the parameter constraints.  With this model, the data at
$\ell > 30$ can be used to predict the low $\ell$ behavior
sufficiently well that the low $\ell$ $P(C_{\ell}|\mathbf{d})$
differences are unimportant.  However, when we allow a logarithmic
scale-dependence to the power-law spectral index, the high $\ell$ data
cannot predict the low $\ell$ data as accurately and the
discrepancies at low $\ell$ are important. We find that the evidence
for a running index is weakened when using our improved description of the
likelihood.

That small differences in $P(C_{\ell}|\mathbf{d})$ can lead to
significant differences in parameter constraints has been pointed out
already by \citet{slosar04}.  They also used a hybrid procedure,
calculating the $\ell \le 12$ likelihood of the parameters directly
from a coarsened version of the \emph{WMAP} maps at every step in the
generation of the Markov chain.  They further used
a more conservative treatment of the uncertainty from foreground
contamination than was used in our and the \emph{WMAP} team's own
modeling. Nevertheless, \citet{slosar04} also find significantly
weakened evidence for non-zero running, in agreement with the present
analysis.

Our current inability to use Gibbs sampling for parameter estimation
over the whole range of $\ell$ (entirely bypassing analytic
approximations to $P(C_{\ell}|\mathbf{d})$) is unfortunate.  With the
inclusion of foregrounds (in particular point sources) and/or with
data from multiple detectors, each with their own beam profile
uncertainties, reliable analytic descriptions of the uncertainty in
$C_{\ell}$ at high $\ell$ do not exist either.  In principle, sampling
approaches can take these uncertainties into account with arbitrary
accuracy.  Below we discuss challenges to extending sampling
procedures to high $\ell$.  Further, we demonstrate that a simple
modification to the BR estimator can dramatically reduce the number of
independent samples required for convergence.

The BR estimate, given samples of maps for temperature and
polarization as well, can easily be extended to estimate the joint
probability distribution of the temperature and polarization auto- and
cross-correlation power spectra.  In contrast, there are no other
existing methods for describing this probability distribution other
than expensive brute-force direct evaluation from the maps, or neglect
of the cross-correlations in the power spectrum errors.  Neglecting
these correlations can lead to significant errors \cite{dore04}.

A strong case for a hybrid estimator, similar to the one used in the
current paper, was made by \citet{efstathiou04}.  The idea was to use
an approximate, but computationally cheap, pseudo-$C_\ell$ method at
high $\ell$, and a more accurate quadratic estimator at low $\ell$'s
where the pseudo-$C_\ell$ approach is significantly sub-optimal.  Here
we point out that Gibbs sampling together with the BR
estimator can replace the quadratic estimator for the low $\ell$
range.  Certainly, the computational cost is significantly higher
because of the sampling stage, and the method does not lend itself as
easily to Monte Carlo simulations.  But BR does have several
advantages.  First, the complete description of uncertainties due to
monopole and dipole subtraction, foreground marginalization and
correlated noise is much more transparent in this approach.  Second,
the computational scalings of the two methods are very different,
implying that the ``low'' $\ell$ regime can be extended to
significantly higher multipoles with the Gibbs sampling method than
with the quadratic estimator.  Finally, the BR estimator accurately
describes the significantly non-Gaussian distribution,
$P(C_\ell|\mathbf{d})$, which is assumed to be Gaussian in
\citep{efstathiou04}.

In section II we briefly review Gibbs sampling and the BR estimator.
In section III we discuss convergence.  In Section IV we compare BR
with the analytic approximations of the \emph{WMAP} likelihood code in
a 2-parameter space of amplitude and tilt, demonstrating the
convergence of the chains and our discrepancies with \emph{WMAP}.  In
section V we present the cosmological parameter results. In section VI
we discuss modifications to BR to allow extension to higher $\ell$
values.  In section VII we conclude.

\section{Gibbs Sampling and the Blackwell-Rao Estimator}

The current paper is a natural extension of the work on CMB analysis
through Gibbs sampling initiated by \citet{jewell04} and
\citet{wandelt03}, and applied to the first-year \emph{WMAP} data by
\citet{eriksen04} and \citet{odwyer04}. Here we only briefly review
the conceptual points behind this method, and refer the interested
reader to those papers for full details.

In this paper we focus on the first-year \emph{WMAP} data, in which 
case the observed data may be written in the form
\begin{equation}
\mathbf{d} = \mathbf{As} + \mathbf{n}.
\end{equation}
Here $\mathbf{d}$ is a noisy sky map, $\mathbf{s}$ is the true sky
signal, $\mathbf{A}$ denotes beam convolution, and $\mathbf{n}$ is
instrumental noise. The sky signal is assumed to be Gaussian
distributed with zero mean and a harmonic space covariance matrix
$\mathbf{C}_{\ell m,\ell' m'} = C_{\ell}
\delta_{\ell\ell'}\delta_{mm'}$.  The noise is also assumed to be
Gaussian distributed, with zero mean and a pixel-space covariance
matrix $\mathbf{N}_{ij} =
\sigma_0^2/\sqrt{N_{\textrm{obs}}(i)}\delta_{ij}$ which is perfectly
known.

\subsection{Elementary Gibbs Sampling}

Our goal is to establish the posterior probability distribution
$P(C_{\ell}|\mathbf{d})$. Since all quantities are assumed to be
Gaussian distributed, this can in principle be done by evaluating the
likelihood function (and assuming a prior). However, this brute-force
approach involves determinant evaluation of a mega-pixel covariance
matrix for modern data sets, and is therefore computationally
unfeasible. An alternative approach was suggested by \citet{jewell04}
and \citet{wandelt03}, namely to draw samples from the posterior,
rather than evaluate it.

While it is difficult to sample from $P(C_{\ell}|\mathbf{d})$
directly, it is in fact fairly straightforward to sample from the
\emph{joint} probability distribution $P(C_{\ell},
\mathbf{s}|\mathbf{d})$ using a method called Gibbs sampling
\citep{gelfand90,tanner96}: Suppose we can sample from the conditional
distributions $P(C_{\ell} | \mathbf{s}, \mathbf{d})$ and $P(\mathbf{s}
| C_{\ell}, \mathbf{d})$. Then the theory of Gibbs sampling says that
samples $(\mathbf{s}^i, C_{\ell}^i)$ can be drawn from the joint
distribution $P(C_{\ell}, \mathbf{s}|\mathbf{d})$ by iterating the
following sampling equations,
\begin{align}
\mathbf{s}^{i+1} &\leftarrow P(\mathbf{s}|C_{\ell}^{i}, \mathbf{d}), \\
C_{\ell}^{i+1} &\leftarrow P(C_{\ell}|\mathbf{s}^{i+1}).
\end{align}
The symbol '$\leftarrow$' indicates that a random vector is drawn from
the distribution on the right hand side. After some burn-in period,
the samples will converge to being drawn from the required joint
distribution. Finally, $P(C_{\ell}|\mathbf{d})$ is found by
marginalizing over $\mathbf{s}$.

How to sample from the required conditional densities is detailed by
\citet{jewell04}, \citet{wandelt03} and \citet{eriksen04}. These
papers also describe how to analyze multi-frequency data, as well as
how to deal with complicating issues such as partial sky coverage and
monopole and dipole contributions. It is also straightforward to
include several forms of foreground marginalization within this
framework, and the uncertainties introduced by any such effects are
naturally expressed by the properties of the sample chains; no
explicit post-processing is required.

\subsection{Parameter Estimation and the Blackwell-Rao Approximation}

By `parameter estimation' we mean mapping out the posterior
distribution $P(\theta|\mathbf{d})$, where $\theta$ is the desired set
of parameters. This is usually done by first choosing some set of
parameters from which a corresponding power spectrum is computed by
numerical codes such as CMBFast.  Second, the distribution value for
the chosen parameters are then found by estimating
$P(C_{\ell}(\theta)|\mathbf{d})$.  This procedure is then either
repeated over a grid in the parameters, or incorporated into an MCMC
chain.

\begin{figure}
\begin{center}
  \mbox{\epsfig{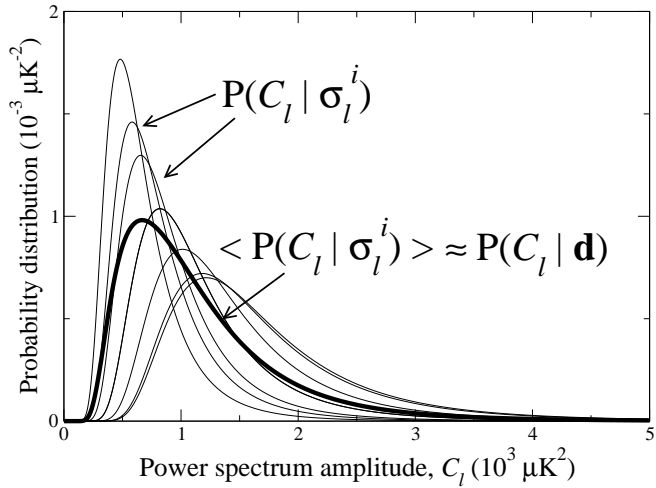}}
  \caption{A one-dimensional illustration of the BR
    estimator. The thin lines indicate the
    $P(C_{\ell}|\sigma_{\ell}^i)$ distributions, and the thick line
    shows their average. This average converges toward the true
    density $P(C_{\ell}|\mathbf{d})$ as the number of samples
    increases.  }
\label{fig:br}
\end{center}
\end{figure}

Thus to estimate parameters we must be able to evaluate
$P(C_{\ell}|\mathbf{d})$ for any model $C_{\ell}$. While we could
compute the histogram of the Gibbs $C_{\ell}$ samples and simply read
off the appropriate values, the BR estimator suggested for this
purpose by \citet{wandelt03} converges more rapidly.  First we expand
the signal sample $\mathbf{s}$ in terms of spherical harmonics,
\begin{equation}
s(\theta,\phi) = \sum_{\ell=0}^{\infty}
\sum_{m=-\ell}^{\ell} s_{\ell m} Y_{\ell m}(\theta, \phi),
\end{equation}
and define its realization-specific power spectrum $\sigma_{\ell}$ by
\begin{equation}
\sigma_{\ell} \equiv
\frac{1}{2\ell+1}\sum_{m=-\ell}^{\ell} |s_{\ell m}|^2.
\end{equation}
Next we note that
\begin{equation}
P(C_{\ell}(\theta)|\mathbf{s},\mathbf{d}) = P(C_{\ell}(\theta)|\mathbf{s}),
\end{equation}
since the power spectrum only depends on the data through the signal
component. Furthermore, it only depends on the signal through
$\sigma_{\ell}$, and not its phases, and therefore
\begin{equation}
P(C_{\ell}(\theta)|\mathbf{s}) = P(C_{\ell}(\theta)|\sigma_{\ell}).
\end{equation}
We may then write \begin{align}
P(C_{\ell}|\mathbf{d}) &= \int P(C_{\ell}, \mathbf{s}|\mathbf{d})
\,\textrm{d}\mathbf{s}\\
&= \int P(C_{\ell}|\mathbf{s},\mathbf{d})
P(\mathbf{s}|\mathbf{d})\,\textrm{d}\mathbf{s} \\
&= \int P(C_{\ell}|\sigma_{\ell})
P(\sigma_{\ell}|\mathbf{d})\,\textrm{D}\sigma_{\ell}\\ &
\approx \frac{1}{N_{\textrm{G}}}\sum_{i=1}^{N_{\textrm{G}}}
P(C_{\ell}|\sigma_{\ell}^{i}),
\label{eq:br}
\end{align}
where $N_{\textrm{G}}$ is the number of Gibbs samples in the
chain. This is called the Blackwell-Rao (BR) estimator for the density
$P(C_{\ell}|\mathbf{d})$. Its meaning is illustrated in Fig.~\ref{fig:br}.

The expression in Eq.~\ref{eq:br} is very useful because, for
a Gaussian field,
\begin{equation}
P(C_{\ell}|\sigma_{\ell})  \propto \prod_{\ell=0}^{\infty} \frac{1}{\sigma_{\ell}}
\left(\frac{\sigma_{\ell}}{C_{\ell}}\right)^{\frac{2\ell+1}{2}}
\, e^{-\frac{2\ell+1}{2} \frac{\sigma_{\ell}}{C_{\ell}}},
\end{equation}
or
\begin{equation}
\ln P(C_{\ell}|\sigma_{\ell}) = \sum_{\ell=0}^{\infty}
\left(\frac{2\ell+1}{2} \left[-\frac{\sigma_{\ell}}{C_{\ell}} +
  \ln\left(\frac{\sigma_{\ell}}{C_{\ell}}\right)\right] - \ln
\sigma_{\ell}\right),
\label{eq:gauss_lnL}
\end{equation}
up to a normalization constant. Eq.~\ref{eq:gauss_lnL} is
straightforward to compute analytically, and an arbitrarily exact
representation of the posterior (with increasing $N_{\textrm{G}}$) may
therefore be established conveniently by means of Eqs.  \ref{eq:br}
and \ref{eq:gauss_lnL}.

\subsection{Comparison with Brute-Force Likelihood Evaluation}

In order to verify that the method works as expected, we apply it to a
simulated map, and compare the results to a brute-force evaluation of
the likelihood. Since this likelihood computation requires inversion
of the signal plus noise covariance matrix, we limit ourselves to a
low-resolution case, with properties similar to those of the
\emph{COBE}-DMR data \citep{bennett96}, but with significantly lower
noise. Specifically, we simulate a sky using the best-fit \emph{WMAP}
power law spectrum, including multipoles between $\ell=2$ and 30. We
then convolve this sky with the DMR beam, add 0.5\% of the 53 GHz DMR
noise (in order to regularize the covariance matrix as the beam drops
off), and finally we apply the extended DMR sky cut.

\begin{figure}
  \begin{center}
    \plotone{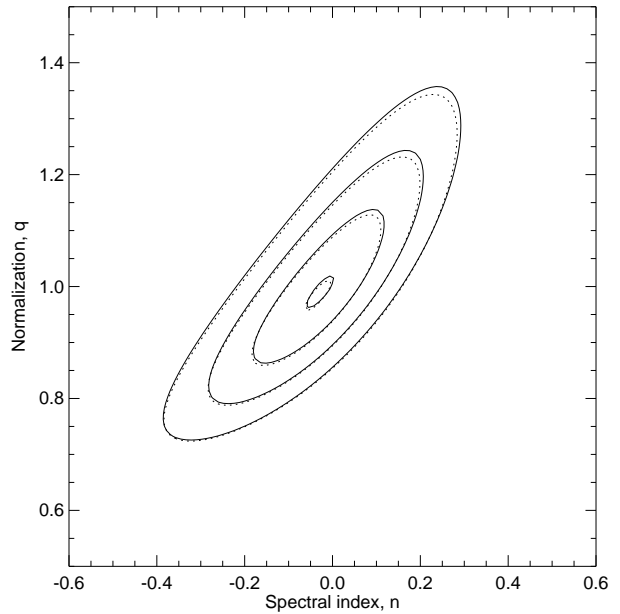}
    \caption{Contours in $(q,n)$ space of constant probability given
the simulated data described in text, for both the BR
estimator (solid) and brute-force evaluation of the likelihood
(dashed). Contours are where $-2\ln P(C_{\ell}|\textbf{d})$ 
rises by 0.1, 2.3, 6.17, and 11.8 from its minimum value,
corresponding (for Gaussian distributions) to the peak, and the 1, 2
and  3$\sigma$ confidence regions.}
\label{fig:br_exact}
\end{center}
\end{figure}

\begin{figure*}

\mbox{\subfigure[Convergence ratio $f=0.06$]{\epsfig{figure=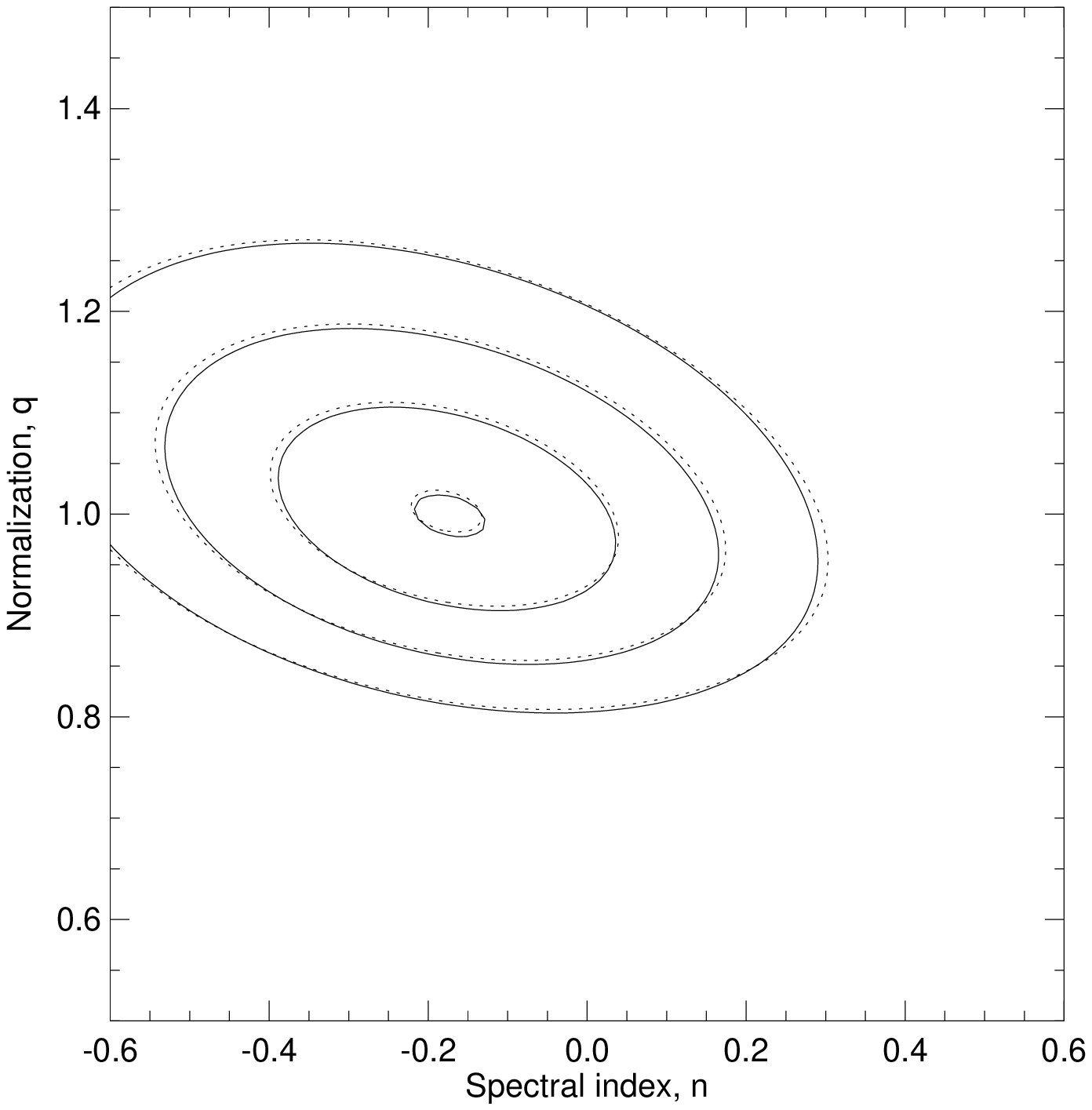,width=0.46\linewidth,clip=}}
      \quad
      \subfigure[Convergence ratio $f=0.47$]{\epsfig{figure=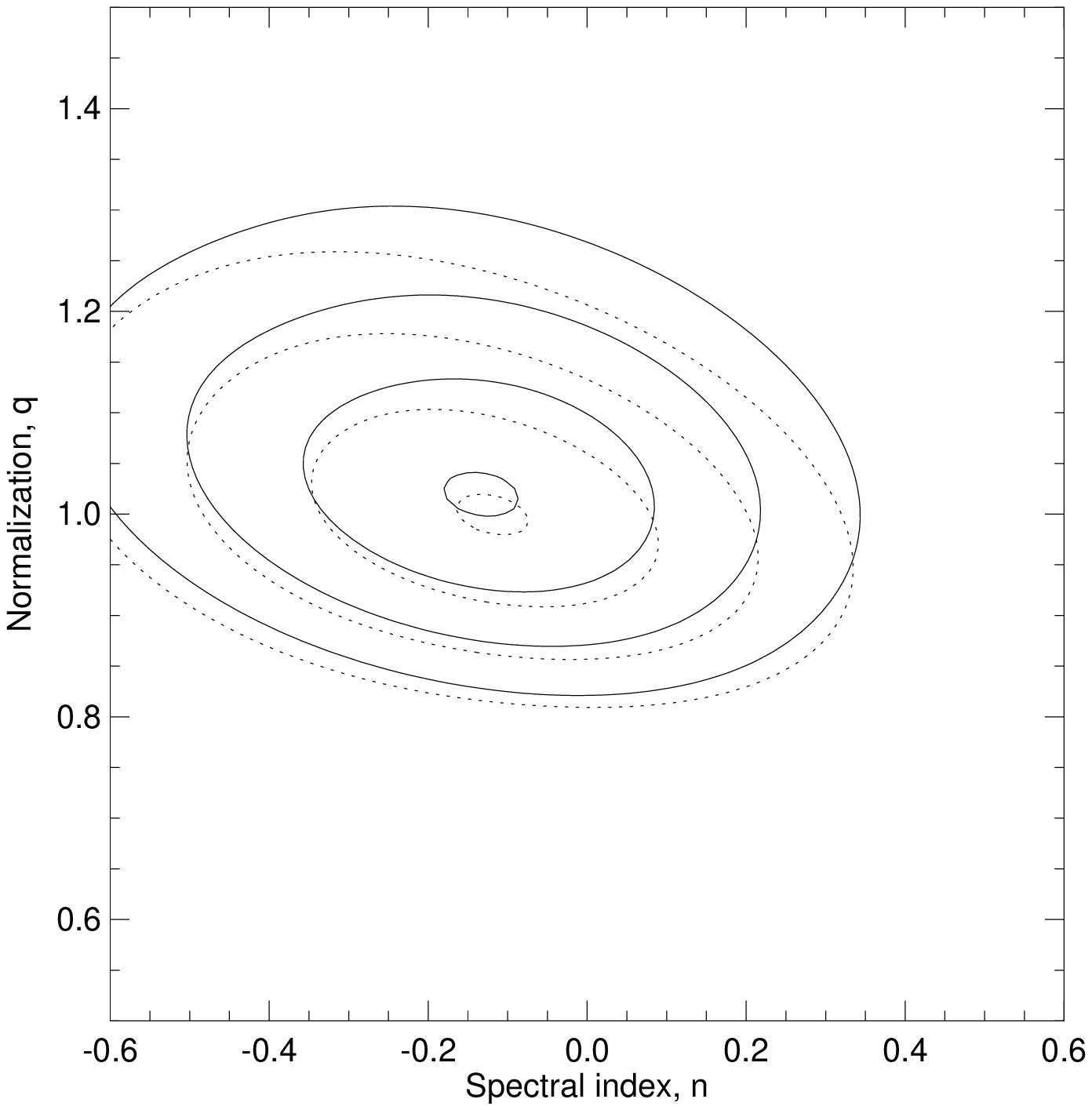,width=0.46\linewidth,clip=}}}

\caption{Illustration of the convergence criterion defined by Eq.
\ref{eq:conv_crit}. If $f \equiv 0$, the two distributions overlap
perfectly, while if $f \equiv 2$, then they are completely
separated. The two distribution pairs shown here have $f = 0.06$ and
$f = 0.47$, respectively. }
\label{fig:comp_two_dist}
\end{figure*}

This simulation is then analyzed both using the Gibbs sampling and
BR machinery as described above, and by computing the full
likelihood over a parameter grid using the Cholesky decomposition
method of \citet{gorski94}. The model power spectrum chosen for
this exercise is of the form
\begin{equation}
\label{eqn:powlaw}
C_{\ell}(q,n) = q \left(\frac{\ell}{\ell_{\textrm{0}}}\right)^{n} C_{\ell}^{\textrm{fid}},
\end{equation}
where $q$ is an amplitude parameter, $n$ is a spectral index,
$\ell_{\textrm{0}}$ is a reference multipole, and
$C_{\ell}^{\textrm{fid}}$ is a fiducial power spectrum, which we take
to be that of a flat $\Lambda$CDM model that fits the data well.  The
fiducial spectrum is chosen to be the input spectrum, and
consequently, we should expect the likelihood of the parameters to
peak near $(q,n) = (1,0)$.

The comparison between the brute-force evaluation and the
BR approximation is not quite as straightforward as one
would like. The problem lies in how to truncate the spherical
harmonics expansion at high $\ell$'s.  The brute-force likelihood
computation requires that the full signal component is contained in
the included harmonic expansion, which means that the noise power has
to be larger than the convolved signal power before truncation.  On
the other hand, the Gibbs sampling approach requires a large number of
samples to converge in this low signal to noise regime.  The
simulation was therefore constructed as a compromise: a very small
amount of noise was added to make the covariance matrix well-behaved
at the very highest $\ell$'s included, but not more than
necessary. Still, small differences between the two approaches must be
expected.

The results from this exercise are shown in Fig.~\ref{fig:br_exact}.
The contours show the lines of constant likelihood where $-2\ln
P(C_{\ell}|\textbf{d})$ rises by 0.1, 2.3, 6.17 and 11.8 from the
minimum, corresponding to the peak and the 1, 2, and $3\sigma$ regions
for a Gaussian distribution. The solid lines show the results from the
BR computation, and the dashed lines show the results from
the exact likelihood computation. Obviously, the agreement between the
two distributions is excellent, considering the very different
approaches taken, and the above-mentioned high-$\ell$ truncation
problem.

\section{Convergence of the BR estimator applied to \emph{WMAP} data}
\label{sec:conv_test}

The ultimate goal of this paper is to apply the methods described
above to the first-year \emph{WMAP} data. In order to do so, we first
need to determine the accuracy of the BR estimator given our finite
number of samples.  In this section we do so by examining how the BR
estimator fluctuates as different subsets of the Gibbs chains are
used.

The Gibbs machinery was applied to the first-year \emph{WMAP} data by
\citet{odwyer04}, and the primary results from that analysis were a
number of $C_{\ell}$ and $\sigma_{\ell}$ sample chains. These chains
are available to us, and form the basis of the following analysis. The
data we use here are those computed from the eight cosmologically
interesting \emph{WMAP} Q-, V- and W-bands, comprising 12 independent
chains of about 80 samples each for a total of 955 samples. For more
details on how these samples were generated, we refer the reader to
\citet{odwyer04} and \citet{eriksen04}.


The main question we need to answer before proceeding with the actual
analysis is, how well does this finite number of samples describe the
full likelihood for a given range of multipoles? To answer this
question, we define a simple test based on the $(q, n)$ model of
Eq.~\ref{eqn:powlaw} as follows: We construct two subsets from
the 955 available samples, each containing $N_{\textrm{s}} < 955/2$
samples, and map out the probability distribution for each subset,
including only multipoles in the range $2 \le \ell \le
\ell_{\textrm{max}}$. From the two resulting probability
distributions, $P_1(q,n)$ and $P_2(q,n)$, we compute the quantity
\begin{equation}
f = \frac{\int|P_1(q,n)-P_2(q,n)| \,\textrm{d}q \textrm{d}n}{\int P_1(q,n) \,\textrm{d}q \textrm{d}n},
\label{eq:conv_crit}
\end{equation}
which measures the relative normalized difference between the two
distributions; if $f \equiv 0$ then the two distributions overlap
perfectly, and if $f \equiv 2$, they are completely separated. We then
increase $N_{\textrm{s}}$ until $f<0.05$ for the first time. Two sets
of such distributions are shown in Fig.~\ref{fig:comp_two_dist},
having $f=0.06$ and $f=0.47$ respectively.

Of course, the chain is likely to go in and out of convergence as
$N_{\textrm{s}}$ is increased further for quite some time, and
therefore there will be a large random contribution to this particular
statistic. For that reason we repeat the experiment eleven times, each
time scrambling the full 955 sample chain, and define the median of
the resulting $N_{\textrm{s}}^{i}$'s as the number of samples required
for convergence\footnote{The rather small number of available samples
prohibits us from studying the asymptotic convergence to a very high
number of samples, and therefore this alternative approach was
chosen.}. The process is then repeated for various values of
$\ell_{\textrm{max}}$.

\begin{figure}

\mbox{\epsfig{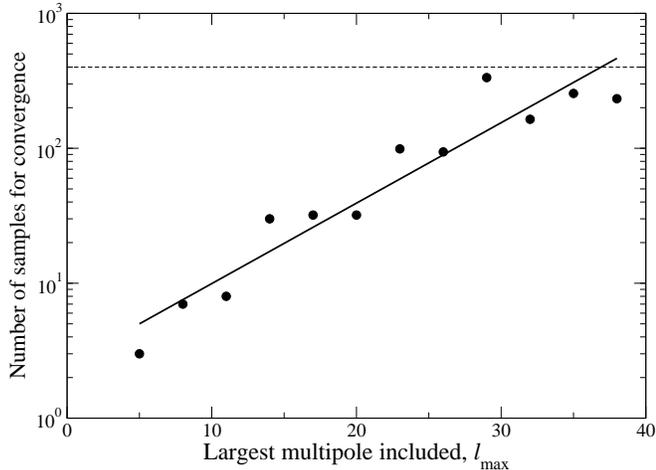}}

\caption{The number of samples required for convergence in the
  BR estimator of the first-year \emph{WMAP} data, as defined in
  Section \ref{sec:conv_test}. The dots indicate the results computed
  from the data as described in the text, while the solid line
  indicates a exponential best-fit. The dashed lines indicate the
  limit above which we cannot probe because of the limited number of
  available samples.  For this calculation we chose $f=0.05$.}
\label{fig:convergence_test}
\end{figure}

The results from this exercise are summarized in
Fig.~\ref{fig:convergence_test}. Two important conclusions may be
drawn from the information shown in this plot. First, the number of
samples required for convergence increases very rapidly with
$\ell_{\textrm{max}}$, possibly following an exponential. However, the
fit is less than perfect, and the true function may possibly be
steeper at the very smallest multipoles than the larger ones, which
would be helpful when probing the higher $\ell$-range. Unfortunately,
the limited number of available samples prohibits us from determining
this function further.

Second, while it is strikingly clear from
Fig.~\ref{fig:convergence_test} that the existing number of samples is
inadequate for probing the full multipole range properly, we may still
conclude that the multipole range $2 \le \ell \le 30$ is quite stable
given that we have 955 samples. In the analysis described in the next
section, we therefore construct a hybrid likelihood consisting of the
BR likelihood for $\ell \le 30$ and the analytic
\emph{WMAP} approximation at higher $\ell$'s.

\begin{figure}
  \begin{center}
    \plotone{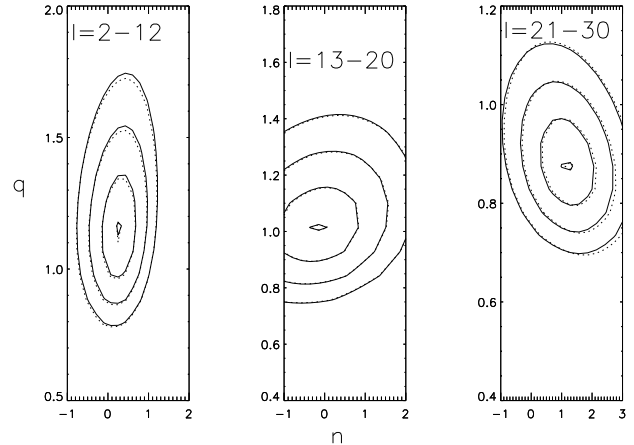}
    \caption{Constraints on $q$ and $n$ where $C_{\ell}(q,n) =
q\,(\ell/\ell_0)^n\,C_{\ell}^{\textrm{fid}}$ and $C_{\ell}^{\textrm{fid}}$
is a fiducial $\Lambda$CDM power spectrum for $\ell_0 = 8$, 17, and 25 from left to right.
Solid lines are for one half of the BR samples and dashed lines are
for another half. Contour levels are as in Fig.~\ref{fig:br_exact}.  }
\label{fig:convergence}
\end{center}
\end{figure}

A second demonstration of the same result is shown in
Fig.~\ref{fig:convergence}, where we have computed the two-parameter
likelihood using the BR estimator, splitting the sample
chain into two parts, for three disjoint $\ell$-ranges ($\ell \in
[2,12], [13,20]$ and $[21,30]$). Here we see that the
estimator is very stable over each of the three $\ell$ ranges.

We have also considered the question of burn-in of the 12 independent
sample chains, by repeating the analyses described above with reduced
chains. Specifically, we removed the five or ten first samples from
each chain, and repeated the analyses. Neither result changed as an
effect of this trimming, implying that burn-in is not a problem for
the Gibbs sampling approach at low $\ell$'s when the estimated
\emph{WMAP} spectrum is used to initialize the Gibbs chains. This
result is in good agreement with the results presented by
\citet{eriksen04}, who showed that the correlation length of the Gibbs
chain is virtually zero when the signal-to-noise is much larger than
one.

\section{Blackwell-Rao vs. {\it WMAP} $P(C_{\ell}|\mathbf{d})$}

\begin{figure*}

\mbox{\epsfig{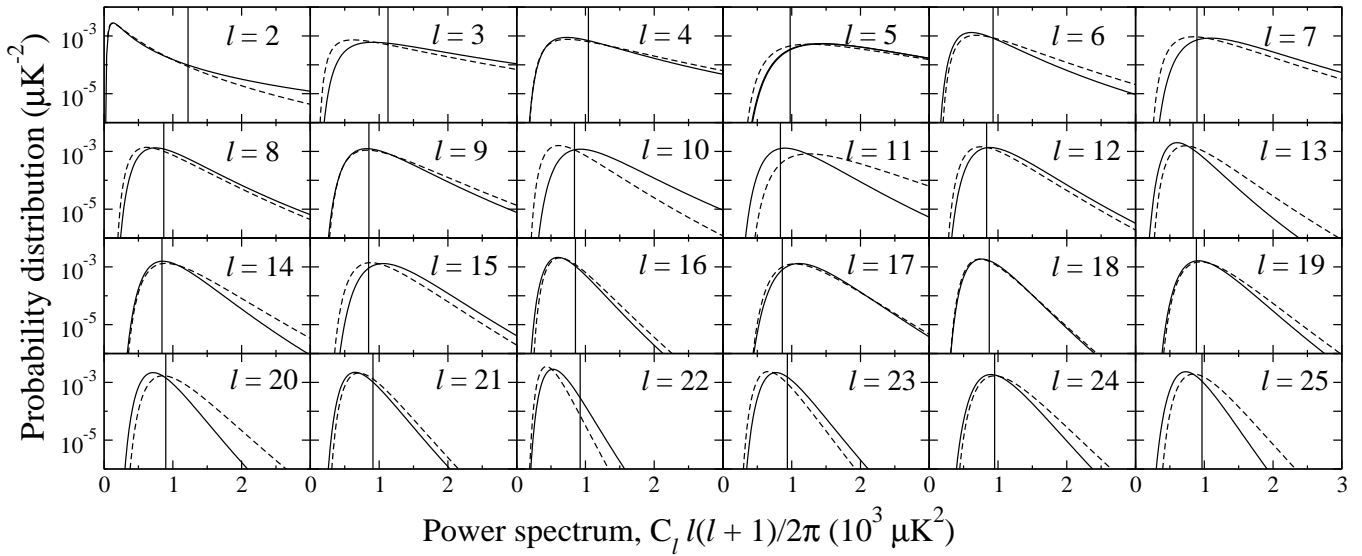}}

\caption{Comparison of the BR (solid curve) and the
  analytic \emph{WMAP} (dashed curve) univariate likelihood functions for each
  multipole up to $\ell = 25$. The vertical lines indicates the value
  of the best-fit \emph{WMAP} power-law model (not including a running
  spectral index). The univariate likelihood functions are computed by
  slicing through the multivariate likelihood, fixing all other
  multipoles at the corresponding best-fit value. Notice that all
  distributions shown here are strongly non-Gaussian.}
\label{fig:wmap_vs_br}
\end{figure*}

There are a number of differences between our analysis and the
\emph{WMAP} team's analysis.  Here we examine the resulting
differences in $P(C_{\ell}|\mathbf{d})$ and in the next section on
estimates of cosmological parameters.  Our goal is to understand the
significance of these low $\ell$ differences.  We do not attempt to
completely disentangle which $P(C_{\ell}|\mathbf{d})$ differences are
due to which analysis differences.

There are at least four areas where the \emph{WMAP} team's analysis 
differs from ours:  
\begin{enumerate}
\item  They use a pseudo-$C_{\ell}$ technique to estimate the
most likely $C_{\ell}$;
\item At $\ell < 100$, in order to reduce residual foreground contamination
they do not include Q-band data;
\item Their pseudo-$C_{\ell}$ estimate
places zero weight on the auto-correlation of maps from the individual 
differencing assemblies; and,
\item They use a variant of the analytic
approximation of \citet{bond00} to the shapes of the likelihoods.
\end{enumerate}

A number of these differences in analysis procedures were discussed by
H03 and \citet{verde03}.  Regarding item 1, one can see in H03 Fig.~12
differences at low $\ell$ between a maximum-likelihood analysis and
pseudo-$C_{\ell}$ analysis as applied to V-band data.  Regarding item
2, in Fig.~3 of H03 one can see differences at low $\ell$ between
inclusion and exclusion of the Q-band data.  Regarding item 3, one can
see differences at low $\ell$ in Fig. 6 of H03 depending on whether
the auto-correlations are included.

\begin{figure}
  \begin{center}
    \mbox{\epsfig{figure=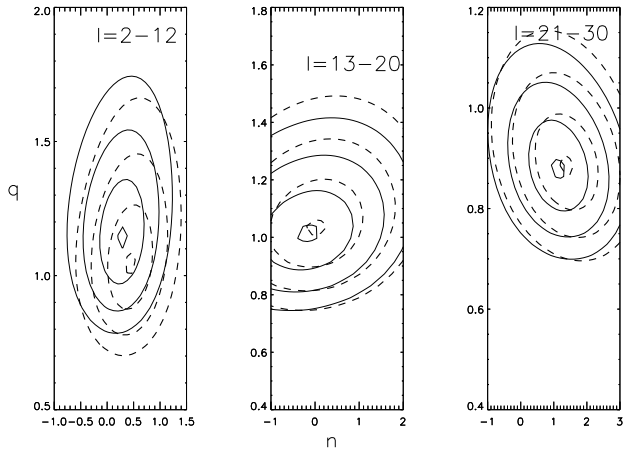,width=\linewidth,clip=}}
    \caption{Constraints on $q$ and $n$ where $C_{\ell}(q,n) =
q\,(\ell/\ell_0)^n\,C_{\ell}^{\textrm{fid}}$ and
$C_{\ell}^{\textrm{fid}}$ is a fiducial $\Lambda$CDM power spectrum 
for $\ell=2$ to
$\ell=12$ with $\ell_0 = 8$ (left panel), $\ell=13$ to $\ell=20$ and
$\ell_0 = 17$ (center panel) and $\ell=21$ to $\ell=30$ with $\ell_0 =
25$ (right panel).  Solid lines are for BR and dashed lines are for
the \emph{WMAP} likelihood code.  Contour levels are as in Figure \ref{fig:br_exact}.  }
\label{fig:amp_tilt2}
\end{center}
\end{figure}

The net result of all these effects is shown in
Fig.~\ref{fig:wmap_vs_br} and \ref{fig:amp_tilt2}.  In
Fig.~\ref{fig:wmap_vs_br}, we compare the univariate likelihood
functions for all multipoles up to $\ell = 25$, as computed using both
the \emph{WMAP} analytic approximation (dashed smooth line) and the
BR approximation (solid smooth line). The BR
likelihoods are computed by varying one single multipole at a time,
keeping the other multipoles fixed at the best-fit power-law model
value.

There are a few clear differences between the two sets of
distributions shown in Fig.~\ref{fig:wmap_vs_br}, the most prominent
being a small horizontal shift in most cases, or in other words,
different power spectrum estimates. This was anticipated, given the
differences discussed above.

More important than these shifts are the relative shapes of the two
distributions. Such features are most easily compared when the two
distributions have identical modes, which is the case for $\ell = 2,
4, 9$ and 14. In the quadrupole case we see that the BR
distribution has a heavier tail than the \emph{WMAP} distribution, while the
opposite is true for the other three cases. On the other hand, we find
spectacular agreement for the $\ell = 17$ and 18 cases. All in all,
the results shown in this figure are consistent with the idea that the
Gibbs sampling approach is an optimal method, while
the \emph{WMAP} approach is based on a pseudo-$C_{\ell}$ method, and
the latter is therefore expected to have slightly larger error
bars. The only case for which this rule is obviously broken is the
quadrupole, and thus we have reason to question the accuracy of this
particular multipole. 

We also note that a similar analysis was carried out by
\citet{slosar04}. One of their major results was a significantly
broader likelihood than the \emph{WMAP} likelihood (as well as a
strong shift toward larger amplitudes) for $\ell \le 10$. The main
difference between that analysis and the present is that they
marginalized over three foreground templates, and, given the results
shown in this section, this additional degree of freedom is most
likely the cause of the broadened likelihood, rather than inherently
under-estimated errors in the \emph{WMAP} likelihood code.
\citet{slosar04} also found a coherent shift toward larger amplitudes.
We see this ourselves to a lesser degree in Fig.~\ref{fig:wmap_vs_br}.
Seven out of the eleven $C_\ell$ in the $\ell = 2$ to 12 range show
some amount of shift to higher $\ell$.

To further study the differences in $P(C_{\ell}|\mathbf{d})$ between
the BR approximation and the analytical approximation used
by the \emph{WMAP} team, we once again adopt the two-parameter
non-physical model defined in Equation \ref{eqn:powlaw},
$C_{\ell}(q,n) = q \left(\ell/\ell_0\right)^n
C_{\ell}^{\textrm{fid}}$, and map out in Fig.~\ref{fig:amp_tilt2} 
the two likelihoods in $q$ and
$n$ using the two approximations.  We display these likelihoods over
the same $\ell$ ranges as in Fig.~\ref{fig:convergence}.  We can see in
the left panel (the $\ell = 2$ to 12 range) clear evidence for a shift
to higher power hinted at by the individual multipole distributions in
Fig.~\ref{fig:wmap_vs_br}.  The peak shifts by $\sim 3/4 \sigma$ and
the BR contours are slightly tighter than the \emph{WMAP} ones.
Discrepancies are smaller in the $\ell = 13$ to 20 range and smaller
still in the 21 to 30 range, especially near $n = 0$.  Note that the
likelihood at $|n| \ga 1$ is irrelevant for physical models since their
spectral shapes do not deviate that strongly from that of the fiducial.

To summarize this section, we have seen that the BR
estimate and that of \emph{WMAP} for $P(C_{\ell} | \mathbf{d})$ do
differ slightly at low $\ell$'s.  This should result in differences in
parameter estimates, to which we now turn.

\section{Effect on cosmological parameters}

We now explore how significant these low $\ell$ differences are for
estimates of cosmological parameters.  We consider two different
cosmological model parameter spaces.  The first is a flat $\Lambda$CDM
cosmology with a power-law primordial power spectrum.  The second
parameter space allows for a logarithmic scale-dependence to the
power-law spectral index so that $n_{\textrm{s}}(k) =
n_{\textrm{s}}(k_0) + \textrm{d}n_{\textrm{s}}/\textrm{d}\ln k
\ln(k/k_0)$.  The $\textrm{d}n_{\textrm{s}}/\textrm{d}\ln k$ parameter
is commonly referred to as the `running of the spectral index', a
reference to the analogous dependence of gauge coupling strength with
energy scale in quantum field theories.

We explore the parameter spaces via the MCMC mechanism as described
by, e.g., \citet{christensen01}.  For the 6-parameter cosmological
models we use $\omega_{\textrm{b}}$, $\omega_{\textrm{d}}$,
$\Omega_{\Lambda}$, $z_{\textrm{rei}}$, $A$, $n_{\textrm{s}}$ (baryon
density, cold dark matter density, dark energy density, redshift of
reionization, amplitude of the primordial power spectrum at $k=0.05
\Mpc^{-1}$ and the scalar index; the total matter density is
$\omega_{\textrm{m}} = \omega_{\textrm{b}} + \omega_{\textrm{d}}$) and
a calibration parameter for each of CBI \citep{pearson03} and ACBAR
\citep{dickinson04,rebolo04}.  For the 7-parameter cosmological model
we use the the same six parameters plus
$\textrm{d}n_{\textrm{s}}/\textrm{d}\ln k$.  We evaluate the
likelihood given the \emph{WMAP} data with the subroutine available at
the LAMBDA\footnote{ The Legacy Archive for Microwave Background Data
Analysis can be found at http://lambda.gsfc.nasa.gov/.} data archive.
For CBI and ACBAR we use the offset log-normal approximation of the
likelihood \citep{bond00}.  The likelihood given all these data
together (referred to as the WMAPext dataset by \citet{spergel03}) is
given by the product of the individual likelihoods.  For the the
hybrid schemes, we replace the the \emph{WMAP} likelihood calculation
for the temperature power spectrum in the range $2 \le \ell \le 30$
with the BR estimator.  In all cases, we employ a prior that is zero
except for models with $0.40 < h < 0.95$, $\tau < 0.30$, and $6.0 <
z_{\textrm{rei}}$ \citep{fan02} in which case the prior is unity.  All
chains have 100,000 samples.

\begin{tablehere}
\begin{table*}[hbt]\small
\begin{center}
\begin{tabular}{c|c|c|c|c|c|c}
\hline
\tableskip
\hline
 & \multicolumn{3}{c}{ $\textrm{d}n_{\textrm{s}}/\textrm{d}\ln k$ fixed to zero} &
 \multicolumn{3}{c}{$\textrm{d}n_{\textrm{s}}/\textrm{d}\ln k$ free to vary} \\ 
\tableskip
Parameter & \emph{WMAP} $P(C_{\ell}|\mathbf{d})$ & hybrid  & difference/$\sigma$ & \emph{WMAP} $P(C_{\ell}|\mathbf{d})$ & hybrid & difference/$\sigma$\\
\tableskip\hline\tableskip
$n_{\textrm{s}}$       & $0.97 \pm 0.03$            &  $0.97 \pm 0.03$   & 0.0 & $0.880 \pm 0.048$   & $0.903\pm 0.047$ & 0.4 \\ 
$\tau$      & $0.132^{+0.097}_{-0.048}$  & $0.140^{+0.080}_{-0.053}$ & 0.1  & $0.202 \pm 0.065$   & $0.208\pm 0.063$ & 0.1  \\ 
$A$         &   $0.80 \pm 0.10$          & $0.79 \pm 0.10$   & 0.1         & $0.91 \pm 0.11$       & $0.90\pm 0.11$ & 0.1   \\ 
$\omega_{\textrm{b}}$  & $0.023 \pm 0.001$          & $0.023 \pm 0.001$ & 0.0  & $0.0215 \pm 0.0013$   & $0.0219 \pm 0.0012$ & 0.3  \\ 
$\omega_{\textrm{m}}$  & $0.136 \pm 0.014$          &  $0.132 \pm 0.013$ & 0.3   & $0.140 \pm .015$    & $0.134 \pm 0.014$  & 0.4 \\ 
$h$         & $0.72 \pm 0.05$            & $0.73 \pm 0.05$   & 0.2  & $0.682 \pm 0.054$   & $0.708\pm 0.054$  & 0.5 \\
$\textrm{d}n_{\textrm{s}}/\textrm{d}\ln k$ & --- & --- & --- & $-0.079 \pm 0.031$ & $-0.063 \pm 0.031$  & 0.5 \\
\tableskip\hline
\end{tabular}
\end{center}
\caption{Cosmological parameter means and standard deviations derived
from the \emph{WMAP}ext dataset using the \emph{WMAP} likelihood code
(columns 2 and 5) and using our hybrid approach where the \emph{WMAP}
likelihood code for the tempearture angular power spectrum is replaced 
at $\ell \le 30$ with our BR
estimate of $P(C_{\ell}|\mathbf{d})$.  The columns labeled
`difference/$\sigma$' give the difference in the parameter means
divided by the standard deviation of the hybrid method.  Note that the
finite number of chain samples gives rise to an uncertainty in each
mean of $\sim 0.1\sigma$.}
\end{table*}
\end{tablehere}

The results for the 6-parameter case using the \emph{WMAP} likelihood
code (column 2 of the table) reproduce those reported by
\citet{spergel03}.  We see that the hybrid scheme leads to almost no
differences, with any shifts in most likely values smaller than $1/3
\sigma$.  Thus there is only a very weak dependence on the differences
in $P(C_{\ell}|\mathbf{d})$ at low $\ell$. The reason for this is that
with the 6-parameter model the data at high $\ell$ tightly constrain
the range of $C_{\ell}$ values at low $\ell$.

Now we turn to the difference between columns 4 and 5, where the only
difference in their derivation is the treatment of the temperature
power spectrum at $\ell \le 30$.  With the extra freedom in the
7-parameter model, the high $\ell$ data can no longer be extrapolated
to low $\ell$ with as much confidence.  The data at low $\ell$ are
thus more informative than in the six-parameter case and the
differences at low $\ell$ become more important.  Four parameters show
shifts greater than $1/3\sigma$: $n_{\textrm{s}}$,
$\omega_{\textrm{b}}$, $\omega_{\textrm{m}}$ and
$\textrm{d}n_{\textrm{s}}/\textrm{d}\ln k$.  The biggest shift is in
$\textrm{d}n_{\textrm{s}}/\textrm{d}\ln k$.  It reduces a $2.5\sigma$
detection to a 2$\sigma$ detection.

We checked to make sure these shifts are significant, given our
limited number of chain elements.  To do so, we looked at 4 subsamples
of the 7-parameter case chains, each with 25,000 samples, to examine
fluctuations in the subsample mean values of each parameter.  We found
these subsample means to deviate from the total sample means with an
rms of $\sim 0.2\sigma$. We thus estimate the sample variance error
in our sample means to be $\sim 0.1\sigma$.  We also ran a chain
with 100,000 samples with the switch at $\ell = 20$, and found it to
be consistent with the hybrid chain with the switch at $\ell=30$.

The direction of the changes is consistent with
Fig.~\ref{fig:amp_tilt2}.  We see our own analysis has a higher level
of power and lower level of tilt in the $\ell=2-12$ range and is more
restrictive of upward power fluctuations in the $\ell=13-20$ range
compared to the \emph{WMAP} team's analysis.  Thus we want the model
power spectra to be more negatively sloped at low $\ell$.  This is
accomplished by the 0.026 increase in the running which reduces
$n_{\textrm{s}}(k)$ at $k=0.009\, Mpc^{-1}$ (which projects to
$\ell=12$) by 0.11 from its value at $k_0 = 0.05 \,\textrm{Mpc}^{-1}$.

It should be noted that the parameter values are strongly dependent on
the high $\tau$ cut.  In fact we have found that most of the
probability is at $\tau > 0.3$, as has been noticed for \emph{WMAP} +
VSA \citep{rebolo04,slosar04} and for \emph{WMAP}+CBI
\citep{readhead04}.  At these high $\tau$ values, the running tends to
be negative also.  Having high $\tau$ and a negative running though is
a priori unlikely in hierarchical models of structure formation, and
is also disfavored when large-scale structure data is included
\citep{slosar04}.

\section{Extending BR to higher $\ell$'s}

We face two challenges to extending the BR estimator to higher $\ell$
values.  The first is that the greater the range of $\ell$ values, the
greater the volume of parameter space to be explored (in units of the
width of the posterior in each direction) and therefore the larger the
number of samples required.  The second is that as the signal-to-noise
ratio drops below unity, the correlation length of the Gibbs samples,
produced by the algorithms of \citet{wandelt03} and \citet{jewell04},
starts to get very long thereby reducing the effective number of
independent samples.  We do not address this second problem here,
which becomes important around $\ell \sim 350$, except to say that we
are currently implementing potential solutions.

We see evidence of the first problem in
Fig.~\ref{fig:convergence_test}.  Here we discuss two solutions, both
of which rely on the low level of dependencies between the errors in
$C_{\ell}$ at different $\ell$ values.  For the first solution, we
replace the BR estimate with a `band BR' estimate where the averaging
over samples is done in discrete bands of $\ell$ that are then
multiplied together.  Specifically, \bea P(\{C_{\ell}\}|\mathbf{d})
&=& \Pi_B \langle P\left( C_{l_<(B)},C_{l_<(B)+1}...,C_{l_>(B)}
\right. \nonumber \\ &&
\left. |\sigma^i_{l_<(B)},\sigma^i_{l_<(B)+1}...,\sigma^i_{l_>(B)}\right)\rangle
\eea where $\langle ... \rangle$ indicates averaging over samples and
the lower and upper $\ell$ values of each band $B$ are denoted by
$\ell_<(B)$ and $\ell_>(B)$.

The advantage of the band BR estimator is that it reduces the volume
of the space to be explored from one with $\ell_{\rm max} - \ell_{\rm min} +1$
dimensions to a product over spaces with number of dimensions equal
to the width of the bands, greatly reducing the volume in units of the
extent of the posterior.  The approximation here ignores inter-band
dependencies.  Tests though have shown these to be negligibly small for
bands of width 12.

\begin{figure}

\mbox{\epsfig{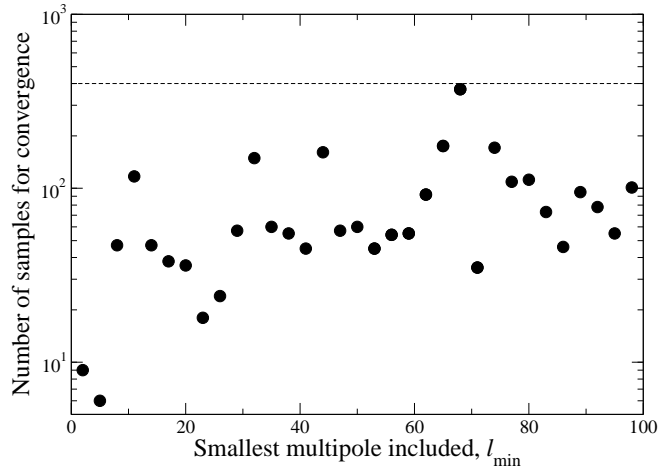}}

\caption{The number of samples required for convergence for the
  band BR estimator of the first-year \emph{WMAP} data,
  as defined in Section \ref{sec:conv_test} for bands extending from
  $\ell_{\rm min}$ to $\ell_{\rm min}+11$. The dots indicate the
  results computed from the data.}
\label{fig:convergence_test2}
\end{figure}

To demonstrate the reduction in the number of samples necessary
for convergence, we re-do Fig.~\ref{fig:convergence_test}.
In Fig.~\ref{fig:convergence_test} $\ell_{\rm min}$ was
fixed to 2 as $\ell_{\rm max}$ increased.  Here as $\ell_{\rm max}$ increases
so does $\ell_{\rm min}$ so that $\ell_{\rm max} - \ell_{\rm min} +1 = 12$.
We see in Fig.~\ref{fig:convergence_test2} that switching to the
band BR estimator flattens out the trend of necessary number of samples
with $\ell_{\rm max}$.

It may be possible to exploit the near-independence of different $\ell$ values
further.  We can use BR (or even a fit to the histogram of $C_{\ell}$ values
in the chain) to estimate univariate marginalized distributions, multiply
these together as if they were independent, and then correct for the
correlations with an analytic correction factor. Namely,
\bea
\ln P(\{C_{\ell}\}|\mathbf{d}) &=& \sum_{\ell} \ln \left(P(C_{\ell}|\mathbf{d})\
\right) \nonumber \\
&+&\sum_{\ell} \frac{\delta C_{\ell}^2}{2 C_{\ell\ell}}
-\sum_{\ell,\ell'} \delta C_{\ell} F_{\ell\ell'} \delta C_{\ell'}/2
\eea
where $\delta C_{\ell} \equiv C_{\ell} - \langle C_{\ell} \rangle$,
$F_{\ell\ell'}$ is the
$C_{\ell}$ Fisher matrix and $C_{\ell\ell'}$ is its inverse.  These
matrices can
be computed as in H03.  Note that the above expression is exact
for a Gaussian distribution, with the first term in the log of the
correction factor simply canceling out the sum of the logs of the
marginalized one-dimensional distributions.
Such a procedure will only require a handful
of independent samples.  Further, one could combine our two solutions
here by using band BR with an analytic correction for the neglected
inter-band dependencies.

Certainly this use of analytics could be extended further to reduce
the demand for number of independent samples.  We expect that an
adequate analytic form can be found for the posterior.  One would then
use the BR estimator, or the $C_{\ell}$ samples, to fit the parameters
of this analytic form.  Such an approach could greatly reduce the
demand for the number of independent samples.  Essentially, we would
be exploiting the fact that $P(\{C_{\ell}\}|\mathbf{d})$ is a very
smooth distribution with a lot of regularity, such as the structure of
inter-$\ell$ correlations and shapes of univariate distributions.
Such an approach will probably be necessary in the high $\ell$ regime
where larger correlation lengths (at least for current sampling
techniques) greatly reduce the number of independent samples.

In the low signal-to-noise regime the number of independent samples required,
even to explore the posterior for a single $\ell$ value, increases because
the width of the BR estimator from an individual sample is much smaller
than the width of the posterior (since the former is for a noiseless sky).
This problem can be mitigated by artificially broadening
the BR kernel.  Specifically, we would set
\bea
\ln\left( P\left(C_{\ell} | \sigma_{\ell}\right)\right) &=& \frac{n_{\ell}}{2}\left[-\frac{\sigma_{\ell}}{C_{\ell}} +
\ln\left(\frac{\sigma_{\ell}}{C_{\ell}}\right)\right] - \ln(\sigma_{\ell}) \nonumber \\
{\rm and \ \ } n_{\ell} & \equiv & (2\ell+1)\left(1+\alpha N_{\ell}/C_{\ell}\right)^{-2}
\eea
where $N_{\ell}$ is the noise contribution to the power spectrum of the
map.
Setting $\alpha > 0$ broadens the kernel for each sample from $\propto
C_{\ell}$ to $\propto (C_{\ell} + \alpha N_{\ell})$.  Unfortunately it
also broadens the posterior from $ \propto (C_{\ell} + N_{\ell})$ to
$\propto \left(C_{\ell} + (1+\alpha) N_{\ell}\right)$.  Thus one
must choose $\alpha$ small enough so the posterior is not overly
broadened. At high $C_{\ell}/N_{\ell}$ this broadening makes no
difference. At low $C_{\ell}/N_{\ell}$ the sample kernel is broadened
by a large factor ($1+ \alpha N_{\ell}/C_{\ell}$) while the posterior is
broadened only by 1+$\alpha$.  Thus one can broaden the sample
kernel in the low signal-to-noise regime (exactly where we want to
broaden it) by a very large amount, without significantly broadening
the posterior. The number of independent samples required for
convergence will drop by this same factor.

Finally, we mention one more way to reduce the dimensionality of the
space to be explored, and thus the number of samples required.  And
that is to replace the $C_{\ell}$'s with band powers.  In the low
signal-to-noise regime such a replacement need not lead to significant
loss of information, assuming models with smooth $C_{\ell}$'s.

\section{Conclusions}

We have found BR to be a useful step in the process of converting CMB
anisotropy data, and a model of it, into estimates of
$P(C_{\ell}|\mathbf{d})$.  We have shown that precise characterization
of this distribution at low $\ell$ is a key step in the estimation of
cosmological parameters.  The differences between
$P(C_{\ell}|\mathbf{d})$ as computed by us with a hybrid approach that
uses BR at $\ell \le 30$ and as computed by the \emph{WMAP} team can lead to
important differences in estimates of cosmological parameters.

The BR estimator converges rapidly at low $\ell$, but requires many
independent samples at high $\ell$.  By exploiting the weak
inter-$\ell$ dependence in $P(C_\ell |\mathbf{d})$ we were able to
modify the BR estimator to greatly improve convergence without
significantly sacrificing accuracy.  Extensions that will allow its
use with correlated data, such as temperature and polarization, or
weak lensing shear from multiple redshift bins, and to higher $\ell$
are worth pursuing.

\begin{acknowledgments}
This work was carried out in the context of the US Planck data
analysis group (USPDC). H. K. E. thanks Dr. Charles R. Lawrence for
arranging his visit to JPL and for great support in general, and also
the Center for Long Wavelength Astrophysics at the Jet Propulsion
Laboratory for its hospitality while this work was completed. H.\ K.\
E.\ acknowledges financial support from the Research Council of
Norway, including a Ph.\ D.\ studentship. We acknowledge use of the
HEALPix software (G\'orski, Hivon \& Wandelt 1998) and analysis
package for deriving the results in this paper.  We also acknowledge
use of the Legacy Archive for Microwave Background Data Analysis
(LAMBDA). This work was partially performed at the Jet Propulsion
Laboratory, California Institute of Technology, under a contract with
the National Aeronautics and Space Administration. This work was
supported by NASA grant NAG5-11098 at UCD.  BDW acknowledges a Beckman
Fellowship from the Center of Advanced Studies at UIUC. This work was
partially supported by NASA/JPL under subcontract 1236748 at UIUC and
subcontract 1230636 at UCD.
\end{acknowledgments}

\bibliography{/work3/knox/bib/cmb3}

\begin{thebibliography}{30}
\expandafter\ifx\csname natexlab\endcsname\relax\def\natexlab#1{#1}\fi
\expandafter\ifx\csname bibnamefont\endcsname\relax
  \def\bibnamefont#1{#1}\fi
\expandafter\ifx\csname bibfnamefont\endcsname\relax
  \def\bibfnamefont#1{#1}\fi
\expandafter\ifx\csname citenamefont\endcsname\relax
  \def\citenamefont#1{#1}\fi
\expandafter\ifx\csname url\endcsname\relax
  \def\url#1{\texttt{#1}}\fi
\expandafter\ifx\csname urlprefix\endcsname\relax\def\urlprefix{URL }\fi
\providecommand{\bibinfo}[2]{#2}
\providecommand{\eprint}[2][]{\url{#2}}

\bibitem[{\citenamefont{{Spergel}}(1995)}]{spergel95}
\bibinfo{author}{\bibfnamefont{D.~N.} \bibnamefont{{Spergel}}}, in
  \emph{\bibinfo{booktitle}{AIP Conf. Proc. 336: Dark Matter}}
  (\bibinfo{year}{1995}), p. \bibinfo{pages}{457}.

\bibitem[{\citenamefont{{Knox}}(1995)}]{knox95}
\bibinfo{author}{\bibfnamefont{L.}~\bibnamefont{{Knox}}},
  \bibinfo{journal}{\prd} \textbf{\bibinfo{volume}{52}}, \bibinfo{pages}{4307}
  (\bibinfo{year}{1995}).

\bibitem[{\citenamefont{{Jungman} et~al.}(1996)\citenamefont{{Jungman},
  {Kamionkowski}, {Kosowsky}, and {Spergel}}}]{jungman96a}
\bibinfo{author}{\bibfnamefont{G.}~\bibnamefont{{Jungman}}},
  \bibinfo{author}{\bibfnamefont{M.}~\bibnamefont{{Kamionkowski}}},
  \bibinfo{author}{\bibfnamefont{A.}~\bibnamefont{{Kosowsky}}},
  \bibnamefont{and} \bibinfo{author}{\bibfnamefont{D.~N.}
  \bibnamefont{{Spergel}}}, \bibinfo{journal}{\prd}
  \textbf{\bibinfo{volume}{54}}, \bibinfo{pages}{1332} (\bibinfo{year}{1996}).

\bibitem[{\citenamefont{{Kuo} et~al.}(2004)\citenamefont{{Kuo}, {Ade}, {Bock},
  {Cantalupo}, {Daub}, {Goldstein}, {Holzapfel}, {Lange}, {Lueker}, {Newcomb}
  et~al.}}]{kuo04}
\bibinfo{author}{\bibfnamefont{C.~L.} \bibnamefont{{Kuo}}},
  \bibinfo{author}{\bibfnamefont{P.~A.~R.} \bibnamefont{{Ade}}},
  \bibinfo{author}{\bibfnamefont{J.~J.} \bibnamefont{{Bock}}},
  \bibinfo{author}{\bibfnamefont{C.}~\bibnamefont{{Cantalupo}}},
  \bibinfo{author}{\bibfnamefont{M.~D.} \bibnamefont{{Daub}}},
  \bibinfo{author}{\bibfnamefont{J.}~\bibnamefont{{Goldstein}}},
  \bibinfo{author}{\bibfnamefont{W.~L.} \bibnamefont{{Holzapfel}}},
  \bibinfo{author}{\bibfnamefont{A.~E.} \bibnamefont{{Lange}}},
  \bibinfo{author}{\bibfnamefont{M.}~\bibnamefont{{Lueker}}},
  \bibinfo{author}{\bibfnamefont{M.}~\bibnamefont{{Newcomb}}},
  \bibnamefont{et~al.}, \bibinfo{journal}{\apj} \textbf{\bibinfo{volume}{600}},
  \bibinfo{pages}{32} (\bibinfo{year}{2004}).

\bibitem[{\citenamefont{{Bennett} et~al.}(2003)\citenamefont{{Bennett},
  {Halpern}, {Hinshaw}, {Jarosik}, {Kogut}, {Limon}, {Meyer}, {Page},
  {Spergel}, {Tucker} et~al.}}]{bennett03}
\bibinfo{author}{\bibfnamefont{C.~L.} \bibnamefont{{Bennett}}},
  \bibinfo{author}{\bibfnamefont{M.}~\bibnamefont{{Halpern}}},
  \bibinfo{author}{\bibfnamefont{G.}~\bibnamefont{{Hinshaw}}},
  \bibinfo{author}{\bibfnamefont{N.}~\bibnamefont{{Jarosik}}},
  \bibinfo{author}{\bibfnamefont{A.}~\bibnamefont{{Kogut}}},
  \bibinfo{author}{\bibfnamefont{M.}~\bibnamefont{{Limon}}},
  \bibinfo{author}{\bibfnamefont{S.~S.} \bibnamefont{{Meyer}}},
  \bibinfo{author}{\bibfnamefont{L.}~\bibnamefont{{Page}}},
  \bibinfo{author}{\bibfnamefont{D.~N.} \bibnamefont{{Spergel}}},
  \bibinfo{author}{\bibfnamefont{G.~S.} \bibnamefont{{Tucker}}},
  \bibnamefont{et~al.}, \bibinfo{journal}{\apjs}
  \textbf{\bibinfo{volume}{148}}, \bibinfo{pages}{1} (\bibinfo{year}{2003}).

\bibitem[{\citenamefont{{Readhead} et~al.}(2004)\citenamefont{{Readhead},
  {Mason}, {Contaldi}, {Pearson}, {Bond}, {Myers}, {Padin}, {Sievers},
  {Cartwright}, {Shepherd} et~al.}}]{readhead04}
\bibinfo{author}{\bibfnamefont{A.~C.~S.} \bibnamefont{{Readhead}}},
  \bibinfo{author}{\bibfnamefont{B.~S.} \bibnamefont{{Mason}}},
  \bibinfo{author}{\bibfnamefont{C.~R.} \bibnamefont{{Contaldi}}},
  \bibinfo{author}{\bibfnamefont{T.~J.} \bibnamefont{{Pearson}}},
  \bibinfo{author}{\bibfnamefont{J.~R.} \bibnamefont{{Bond}}},
  \bibinfo{author}{\bibfnamefont{S.~T.} \bibnamefont{{Myers}}},
  \bibinfo{author}{\bibfnamefont{S.}~\bibnamefont{{Padin}}},
  \bibinfo{author}{\bibfnamefont{J.~L.} \bibnamefont{{Sievers}}},
  \bibinfo{author}{\bibfnamefont{J.~K.} \bibnamefont{{Cartwright}}},
  \bibinfo{author}{\bibfnamefont{M.~C.} \bibnamefont{{Shepherd}}},
  \bibnamefont{et~al.}, \bibinfo{journal}{\apj} \textbf{\bibinfo{volume}{609}},
  \bibinfo{pages}{498} (\bibinfo{year}{2004}).

\bibitem[{\citenamefont{{Spergel} et~al.}(2003)\citenamefont{{Spergel},
  {Verde}, {Peiris}, {Komatsu}, {Nolta}, {Bennett}, {Halpern}, {Hinshaw},
  {Jarosik}, {Kogut} et~al.}}]{spergel03}
\bibinfo{author}{\bibfnamefont{D.~N.} \bibnamefont{{Spergel}}},
  \bibinfo{author}{\bibfnamefont{L.}~\bibnamefont{{Verde}}},
  \bibinfo{author}{\bibfnamefont{H.~V.} \bibnamefont{{Peiris}}},
  \bibinfo{author}{\bibfnamefont{E.}~\bibnamefont{{Komatsu}}},
  \bibinfo{author}{\bibfnamefont{M.~R.} \bibnamefont{{Nolta}}},
  \bibinfo{author}{\bibfnamefont{C.~L.} \bibnamefont{{Bennett}}},
  \bibinfo{author}{\bibfnamefont{M.}~\bibnamefont{{Halpern}}},
  \bibinfo{author}{\bibfnamefont{G.}~\bibnamefont{{Hinshaw}}},
  \bibinfo{author}{\bibfnamefont{N.}~\bibnamefont{{Jarosik}}},
  \bibinfo{author}{\bibfnamefont{A.}~\bibnamefont{{Kogut}}},
  \bibnamefont{et~al.}, \bibinfo{journal}{\apjs}
  \textbf{\bibinfo{volume}{148}}, \bibinfo{pages}{175} (\bibinfo{year}{2003}).

\bibitem[{\citenamefont{{Goldstein} et~al.}(2003)\citenamefont{{Goldstein},
  {Ade}, {Bock}, {Bond}, {Cantalupo}, {Contaldi}, {Daub}, {Holzapfel}, {Kuo},
  {Lange} et~al.}}]{goldstein03}
\bibinfo{author}{\bibfnamefont{J.~H.} \bibnamefont{{Goldstein}}},
  \bibinfo{author}{\bibfnamefont{P.~A.~R.} \bibnamefont{{Ade}}},
  \bibinfo{author}{\bibfnamefont{J.~J.} \bibnamefont{{Bock}}},
  \bibinfo{author}{\bibfnamefont{J.~R.} \bibnamefont{{Bond}}},
  \bibinfo{author}{\bibfnamefont{C.}~\bibnamefont{{Cantalupo}}},
  \bibinfo{author}{\bibfnamefont{C.~R.} \bibnamefont{{Contaldi}}},
  \bibinfo{author}{\bibfnamefont{M.~D.} \bibnamefont{{Daub}}},
  \bibinfo{author}{\bibfnamefont{W.~L.} \bibnamefont{{Holzapfel}}},
  \bibinfo{author}{\bibfnamefont{C.}~\bibnamefont{{Kuo}}},
  \bibinfo{author}{\bibfnamefont{A.~E.} \bibnamefont{{Lange}}},
  \bibnamefont{et~al.}, \bibinfo{journal}{\apj} \textbf{\bibinfo{volume}{599}},
  \bibinfo{pages}{773} (\bibinfo{year}{2003}).

\bibitem[{\citenamefont{{Rebolo} et~al.}(2004)\citenamefont{{Rebolo}, {Battye},
  {Carreira}, {Cleary}, {Davies}, {Davis}, {Dickinson}, {Genova-Santos},
  {Grainge}, {Gutirrez} et~al.}}]{rebolo04}
\bibinfo{author}{\bibfnamefont{R.}~\bibnamefont{{Rebolo}}},
  \bibinfo{author}{\bibfnamefont{R.~A.} \bibnamefont{{Battye}}},
  \bibinfo{author}{\bibfnamefont{P.}~\bibnamefont{{Carreira}}},
  \bibinfo{author}{\bibfnamefont{K.}~\bibnamefont{{Cleary}}},
  \bibinfo{author}{\bibfnamefont{R.~D.} \bibnamefont{{Davies}}},
  \bibinfo{author}{\bibfnamefont{R.~J.} \bibnamefont{{Davis}}},
  \bibinfo{author}{\bibfnamefont{C.}~\bibnamefont{{Dickinson}}},
  \bibinfo{author}{\bibfnamefont{R.}~\bibnamefont{{Genova-Santos}}},
  \bibinfo{author}{\bibfnamefont{K.}~\bibnamefont{{Grainge}}},
  \bibinfo{author}{\bibfnamefont{C.~M.} \bibnamefont{{Gutirrez}}},
  \bibnamefont{et~al.}, \bibinfo{journal}{\mnras, submitted}
  (\bibinfo{year}{2004}), \eprint{astro-ph/0402466}.

\bibitem[{\citenamefont{{Gilks} et~al.}(1996)\citenamefont{{Gilks}, S., and
  {Spiegelhalter}}}]{gilks96}
\bibinfo{author}{\bibfnamefont{W.~R.} \bibnamefont{{Gilks}}},
  \bibinfo{author}{\bibfnamefont{R.}~\bibnamefont{S.}}, \bibnamefont{and}
  \bibinfo{author}{\bibfnamefont{D.~J.} \bibnamefont{{Spiegelhalter}}},
  \emph{\bibinfo{title}{Markov Chain Monte Carlo in Practice}}
  (\bibinfo{publisher}{Chapman and Hall}, \bibinfo{address}{London},
  \bibinfo{year}{1996}).

\bibitem[{\citenamefont{{Christensen} et~al.}(2001)\citenamefont{{Christensen},
  {Meyer}, {Knox}, and {Luey}}}]{christensen01}
\bibinfo{author}{\bibfnamefont{N.}~\bibnamefont{{Christensen}}},
  \bibinfo{author}{\bibfnamefont{R.}~\bibnamefont{{Meyer}}},
  \bibinfo{author}{\bibfnamefont{L.}~\bibnamefont{{Knox}}}, \bibnamefont{and}
  \bibinfo{author}{\bibfnamefont{B.}~\bibnamefont{{Luey}}},
  \bibinfo{journal}{Classical Quantum Gravity} \textbf{\bibinfo{volume}{18}},
  \bibinfo{pages}{2677} (\bibinfo{year}{2001}).

\bibitem[{\citenamefont{{Knox} et~al.}(2001)\citenamefont{{Knox},
  {Christensen}, and {Skordis}}}]{knox01b}
\bibinfo{author}{\bibfnamefont{L.}~\bibnamefont{{Knox}}},
  \bibinfo{author}{\bibfnamefont{N.}~\bibnamefont{{Christensen}}},
  \bibnamefont{and}
  \bibinfo{author}{\bibfnamefont{C.}~\bibnamefont{{Skordis}}},
  \bibinfo{journal}{\apjl} \textbf{\bibinfo{volume}{563}}, \bibinfo{pages}{L95}
  (\bibinfo{year}{2001}).

\bibitem[{\citenamefont{{Verde} et~al.}(2003)\citenamefont{{Verde}, {Peiris},
  {Spergel}, {Nolta}, {Bennett}, {Halpern}, {Hinshaw}, {Jarosik}, {Kogut},
  {Limon} et~al.}}]{verde03}
\bibinfo{author}{\bibfnamefont{L.}~\bibnamefont{{Verde}}},
  \bibinfo{author}{\bibfnamefont{H.~V.} \bibnamefont{{Peiris}}},
  \bibinfo{author}{\bibfnamefont{D.~N.} \bibnamefont{{Spergel}}},
  \bibinfo{author}{\bibfnamefont{M.~R.} \bibnamefont{{Nolta}}},
  \bibinfo{author}{\bibfnamefont{C.~L.} \bibnamefont{{Bennett}}},
  \bibinfo{author}{\bibfnamefont{M.}~\bibnamefont{{Halpern}}},
  \bibinfo{author}{\bibfnamefont{G.}~\bibnamefont{{Hinshaw}}},
  \bibinfo{author}{\bibfnamefont{N.}~\bibnamefont{{Jarosik}}},
  \bibinfo{author}{\bibfnamefont{A.}~\bibnamefont{{Kogut}}},
  \bibinfo{author}{\bibfnamefont{M.}~\bibnamefont{{Limon}}},
  \bibnamefont{et~al.}, \bibinfo{journal}{\apjs}
  \textbf{\bibinfo{volume}{148}}, \bibinfo{pages}{195} (\bibinfo{year}{2003}).

\bibitem[{\citenamefont{{Bond} et~al.}(2000)\citenamefont{{Bond}, {Jaffe}, and
  {Knox}}}]{bond00}
\bibinfo{author}{\bibfnamefont{J.~R.} \bibnamefont{{Bond}}},
  \bibinfo{author}{\bibfnamefont{A.~H.} \bibnamefont{{Jaffe}}},
  \bibnamefont{and} \bibinfo{author}{\bibfnamefont{L.}~\bibnamefont{{Knox}}},
  \bibinfo{journal}{\apj} \textbf{\bibinfo{volume}{533}}, \bibinfo{pages}{19}
  (\bibinfo{year}{2000}).

\bibitem[{\citenamefont{{Bartlett} et~al.}(2000)\citenamefont{{Bartlett},
  {Douspis}, {Blanchard}, and {Le Dour}}}]{bartlett00}
\bibinfo{author}{\bibfnamefont{J.~G.} \bibnamefont{{Bartlett}}},
  \bibinfo{author}{\bibfnamefont{M.}~\bibnamefont{{Douspis}}},
  \bibinfo{author}{\bibfnamefont{A.}~\bibnamefont{{Blanchard}}},
  \bibnamefont{and} \bibinfo{author}{\bibfnamefont{M.}~\bibnamefont{{Le
  Dour}}}, \bibinfo{journal}{\aaps} \textbf{\bibinfo{volume}{146}},
  \bibinfo{pages}{507} (\bibinfo{year}{2000}).

\bibitem[{\citenamefont{{Wandelt} et~al.}(2004)\citenamefont{{Wandelt},
  {Larson}, and {Lakshminarayanan}}}]{wandelt03}
\bibinfo{author}{\bibfnamefont{B.~D.} \bibnamefont{{Wandelt}}},
  \bibinfo{author}{\bibfnamefont{D.~L.} \bibnamefont{{Larson}}},
  \bibnamefont{and}
  \bibinfo{author}{\bibfnamefont{A.}~\bibnamefont{{Lakshminarayanan}}},
  \bibinfo{journal}{\prd} \textbf{\bibinfo{volume}{70}},
  \bibinfo{pages}{083511} (\bibinfo{year}{2004}).

\bibitem[{\citenamefont{{Eriksen} et~al.}(2004)\citenamefont{{Eriksen},
  {O'Dwyer}, {Jewell}, {Wandelt}, {Larson}, {G\'{o}rski}, {Levin}, {Banday},
  and {Lilje}}}]{eriksen04}
\bibinfo{author}{\bibfnamefont{H.~K.} \bibnamefont{{Eriksen}}},
  \bibinfo{author}{\bibfnamefont{I.~J.} \bibnamefont{{O'Dwyer}}},
  \bibinfo{author}{\bibfnamefont{J.~B.} \bibnamefont{{Jewell}}},
  \bibinfo{author}{\bibfnamefont{B.~D.} \bibnamefont{{Wandelt}}},
  \bibinfo{author}{\bibfnamefont{D.~L.} \bibnamefont{{Larson}}},
  \bibinfo{author}{\bibfnamefont{K.~M.} \bibnamefont{{G\'{o}rski}}},
  \bibinfo{author}{\bibfnamefont{S.}~\bibnamefont{{Levin}}},
  \bibinfo{author}{\bibfnamefont{A.~J.} \bibnamefont{{Banday}}},
  \bibnamefont{and} \bibinfo{author}{\bibfnamefont{P.~B.}
  \bibnamefont{{Lilje}}}, \bibinfo{journal}{\apjs}
  \textbf{\bibinfo{volume}{155}}, \bibinfo{pages}{227} (\bibinfo{year}{2004}).

\bibitem[{\citenamefont{{O'Dwyer} et~al.}(2004)\citenamefont{{O'Dwyer},
  {Eriksen}, {Wandelt}, {Jewell}, {Larson}, {G\'{o}rski}, {Levin}, {Banday},
  and {Lilje}}}]{odwyer04}
\bibinfo{author}{\bibfnamefont{I.~J.} \bibnamefont{{O'Dwyer}}},
  \bibinfo{author}{\bibfnamefont{H.~K.} \bibnamefont{{Eriksen}}},
  \bibinfo{author}{\bibfnamefont{B.~D.} \bibnamefont{{Wandelt}}},
  \bibinfo{author}{\bibfnamefont{J.~B.} \bibnamefont{{Jewell}}},
  \bibinfo{author}{\bibfnamefont{D.~L.} \bibnamefont{{Larson}}},
  \bibinfo{author}{\bibfnamefont{K.~M.} \bibnamefont{{G\'{o}rski}}},
  \bibinfo{author}{\bibfnamefont{S.}~\bibnamefont{{Levin}}},
  \bibinfo{author}{\bibfnamefont{A.~J.} \bibnamefont{{Banday}}},
  \bibnamefont{and} \bibinfo{author}{\bibfnamefont{P.~B.}
  \bibnamefont{{Lilje}}}, \bibinfo{journal}{\apjs}
  \textbf{\bibinfo{volume}{612}} (\bibinfo{year}{2004}).

\bibitem[{\citenamefont{{Hinshaw} et~al.}(2003)\citenamefont{{Hinshaw},
  {Spergel}, {Verde}, {Hill}, {Meyer}, {Barnes}, {Bennett}, {Halpern},
  {Jarosik}, {Kogut} et~al.}}]{hinshaw03}
\bibinfo{author}{\bibfnamefont{G.}~\bibnamefont{{Hinshaw}}},
  \bibinfo{author}{\bibfnamefont{D.~N.} \bibnamefont{{Spergel}}},
  \bibinfo{author}{\bibfnamefont{L.}~\bibnamefont{{Verde}}},
  \bibinfo{author}{\bibfnamefont{R.~S.} \bibnamefont{{Hill}}},
  \bibinfo{author}{\bibfnamefont{S.~S.} \bibnamefont{{Meyer}}},
  \bibinfo{author}{\bibfnamefont{C.}~\bibnamefont{{Barnes}}},
  \bibinfo{author}{\bibfnamefont{C.~L.} \bibnamefont{{Bennett}}},
  \bibinfo{author}{\bibfnamefont{M.}~\bibnamefont{{Halpern}}},
  \bibinfo{author}{\bibfnamefont{N.}~\bibnamefont{{Jarosik}}},
  \bibinfo{author}{\bibfnamefont{A.}~\bibnamefont{{Kogut}}},
  \bibnamefont{et~al.}, \bibinfo{journal}{\apjs}
  \textbf{\bibinfo{volume}{148}}, \bibinfo{pages}{135} (\bibinfo{year}{2003}).

\bibitem[{\citenamefont{{Slosar} et~al.}(2004)\citenamefont{{Slosar}, {Seljak},
  and {Makarov}}}]{slosar04}
\bibinfo{author}{\bibfnamefont{A.}~\bibnamefont{{Slosar}}},
  \bibinfo{author}{\bibfnamefont{U.}~\bibnamefont{{Seljak}}}, \bibnamefont{and}
  \bibinfo{author}{\bibfnamefont{A.}~\bibnamefont{{Makarov}}},
  \bibinfo{journal}{\prd} \textbf{\bibinfo{volume}{69}},
  \bibinfo{pages}{123003} (\bibinfo{year}{2004}).

\bibitem[{\citenamefont{{Dor{\' e}} et~al.}(2004)\citenamefont{{Dor{\' e}},
  {Holder}, and {Loeb}}}]{dore04}
\bibinfo{author}{\bibfnamefont{O.}~\bibnamefont{{Dor{\' e}}}},
  \bibinfo{author}{\bibfnamefont{G.~P.} \bibnamefont{{Holder}}},
  \bibnamefont{and} \bibinfo{author}{\bibfnamefont{A.}~\bibnamefont{{Loeb}}},
  \bibinfo{journal}{\apj} \textbf{\bibinfo{volume}{612}}, \bibinfo{pages}{81}
  (\bibinfo{year}{2004}).

\bibitem[{\citenamefont{{Efstathiou}}(2004)}]{efstathiou04}
\bibinfo{author}{\bibfnamefont{G.}~\bibnamefont{{Efstathiou}}},
  \bibinfo{journal}{\mnras} \textbf{\bibinfo{volume}{349}},
  \bibinfo{pages}{603} (\bibinfo{year}{2004}).

\bibitem[{\citenamefont{{Jewell} et~al.}(2004)\citenamefont{{Jewell}, {Levin},
  and {Anderson}}}]{jewell04}
\bibinfo{author}{\bibfnamefont{J.}~\bibnamefont{{Jewell}}},
  \bibinfo{author}{\bibfnamefont{S.}~\bibnamefont{{Levin}}}, \bibnamefont{and}
  \bibinfo{author}{\bibfnamefont{C.~H.} \bibnamefont{{Anderson}}},
  \bibinfo{journal}{\apj} \textbf{\bibinfo{volume}{609}}, \bibinfo{pages}{1}
  (\bibinfo{year}{2004}).

\bibitem[{\citenamefont{{Gelfand} and {Smith}}(1990)}]{gelfand90}
\bibinfo{author}{\bibfnamefont{A.~E.} \bibnamefont{{Gelfand}}}
  \bibnamefont{and} \bibinfo{author}{\bibfnamefont{A.~F.~M.}
  \bibnamefont{{Smith}}}, \bibinfo{journal}{J. Am. Stat. Asso.}
  \textbf{\bibinfo{volume}{85}}, \bibinfo{pages}{398} (\bibinfo{year}{1990}).

\bibitem[{\citenamefont{{Tanner}}(1996)}]{tanner96}
\bibinfo{author}{\bibfnamefont{M.~A.} \bibnamefont{{Tanner}}},
  \emph{\bibinfo{title}{{Tools for statistical inference}}}
  (\bibinfo{publisher}{Springer Verlag, New York}, \bibinfo{year}{1996}).

\bibitem[{\citenamefont{{Bennett} et~al.}(1996)\citenamefont{{Bennett},
  {Banday}, {Gorski}, {Hinshaw}, {Jackson}, {Keegstra}, {Kogut}, {Smoot},
  {Wilkinson}, and {Wright}}}]{bennett96}
\bibinfo{author}{\bibfnamefont{C.~L.} \bibnamefont{{Bennett}}},
  \bibinfo{author}{\bibfnamefont{A.~J.} \bibnamefont{{Banday}}},
  \bibinfo{author}{\bibfnamefont{K.~M.} \bibnamefont{{Gorski}}},
  \bibinfo{author}{\bibfnamefont{G.}~\bibnamefont{{Hinshaw}}},
  \bibinfo{author}{\bibfnamefont{P.}~\bibnamefont{{Jackson}}},
  \bibinfo{author}{\bibfnamefont{P.}~\bibnamefont{{Keegstra}}},
  \bibinfo{author}{\bibfnamefont{A.}~\bibnamefont{{Kogut}}},
  \bibinfo{author}{\bibfnamefont{G.~F.} \bibnamefont{{Smoot}}},
  \bibinfo{author}{\bibfnamefont{D.~T.} \bibnamefont{{Wilkinson}}},
  \bibnamefont{and} \bibinfo{author}{\bibfnamefont{E.~L.}
  \bibnamefont{{Wright}}}, \bibinfo{journal}{\apjl}
  \textbf{\bibinfo{volume}{464}}, \bibinfo{pages}{L1} (\bibinfo{year}{1996}).

\bibitem[{\citenamefont{{G\'{o}rski}}(1994)}]{gorski94}
\bibinfo{author}{\bibfnamefont{K.~M.} \bibnamefont{{G\'{o}rski}}},
  \bibinfo{journal}{\apjl} \textbf{\bibinfo{volume}{430}}, \bibinfo{pages}{L85}
  (\bibinfo{year}{1994}).

\bibitem[{\citenamefont{{Pearson} et~al.}(2003)\citenamefont{{Pearson},
  {Mason}, {Readhead}, {Shepherd}, {Sievers}, {Udomprasert}, {Cartwright},
  {Farmer}, {Padin}, {Myers} et~al.}}]{pearson03}
\bibinfo{author}{\bibfnamefont{T.~J.} \bibnamefont{{Pearson}}},
  \bibinfo{author}{\bibfnamefont{B.~S.} \bibnamefont{{Mason}}},
  \bibinfo{author}{\bibfnamefont{A.~C.~S.} \bibnamefont{{Readhead}}},
  \bibinfo{author}{\bibfnamefont{M.~C.} \bibnamefont{{Shepherd}}},
  \bibinfo{author}{\bibfnamefont{J.~L.} \bibnamefont{{Sievers}}},
  \bibinfo{author}{\bibfnamefont{P.~S.} \bibnamefont{{Udomprasert}}},
  \bibinfo{author}{\bibfnamefont{J.~K.} \bibnamefont{{Cartwright}}},
  \bibinfo{author}{\bibfnamefont{A.~J.} \bibnamefont{{Farmer}}},
  \bibinfo{author}{\bibfnamefont{S.}~\bibnamefont{{Padin}}},
  \bibinfo{author}{\bibfnamefont{S.~T.} \bibnamefont{{Myers}}},
  \bibnamefont{et~al.}, \bibinfo{journal}{\apj} \textbf{\bibinfo{volume}{591}},
  \bibinfo{pages}{556} (\bibinfo{year}{2003}).

\bibitem[{\citenamefont{{Dickinson} et~al.}(2004)\citenamefont{{Dickinson},
  {Battye}, {Cleary}, {Davies}, {Davis}, {Genova-Santos}, {Grainge},
  {Gutierrez}, {Hafez}, {Hobson} et~al.}}]{dickinson04}
\bibinfo{author}{\bibfnamefont{C.}~\bibnamefont{{Dickinson}}},
  \bibinfo{author}{\bibfnamefont{R.~A.} \bibnamefont{{Battye}}},
  \bibinfo{author}{\bibfnamefont{K.}~\bibnamefont{{Cleary}}},
  \bibinfo{author}{\bibfnamefont{R.~D.} \bibnamefont{{Davies}}},
  \bibinfo{author}{\bibfnamefont{R.~J.} \bibnamefont{{Davis}}},
  \bibinfo{author}{\bibfnamefont{R.}~\bibnamefont{{Genova-Santos}}},
  \bibinfo{author}{\bibfnamefont{K.}~\bibnamefont{{Grainge}}},
  \bibinfo{author}{\bibfnamefont{C.~M.} \bibnamefont{{Gutierrez}}},
  \bibinfo{author}{\bibfnamefont{Y.~A.} \bibnamefont{{Hafez}}},
  \bibinfo{author}{\bibfnamefont{M.~P.} \bibnamefont{{Hobson}}},
  \bibnamefont{et~al.}, \bibinfo{journal}{\mnras, submitted}
  (\bibinfo{year}{2004}), \eprint{astro-ph/0402498}.

\bibitem[{\citenamefont{{Fan} et~al.}(2002)\citenamefont{{Fan}, {Narayanan},
  {Strauss}, {White}, {Becker}, {Pentericci}, and {Rix}}}]{fan02}
\bibinfo{author}{\bibfnamefont{X.}~\bibnamefont{{Fan}}},
  \bibinfo{author}{\bibfnamefont{V.~K.} \bibnamefont{{Narayanan}}},
  \bibinfo{author}{\bibfnamefont{M.~A.} \bibnamefont{{Strauss}}},
  \bibinfo{author}{\bibfnamefont{R.~L.} \bibnamefont{{White}}},
  \bibinfo{author}{\bibfnamefont{R.~H.} \bibnamefont{{Becker}}},
  \bibinfo{author}{\bibfnamefont{L.}~\bibnamefont{{Pentericci}}},
  \bibnamefont{and} \bibinfo{author}{\bibfnamefont{H.}~\bibnamefont{{Rix}}},
  \bibinfo{journal}{\aj} \textbf{\bibinfo{volume}{123}}, \bibinfo{pages}{1247}
  (\bibinfo{year}{2002}).

\end{thebibliography}

\end{document}